\documentstyle[preprint,aps]{revtex}

\tightenlines

\begin{document}

\draft

\title{
On the universality of the a priori mixing angles in\\
weak radiative decays of hyperons
}

\author{
A. Garc\'{\i}a
}
\address{
Departamento de F\'{\i}sica.
Centro de Investigaci\'on y de Estudios Avanzados del IPN\\
A.P. 14-740. M\'exico, D.F. 07000. MEXICO\\ 
}
\author{ 
R.~Huerta
}
\address{
Departamento de F\'{\i}sica Aplicada\\
Centro de Investigaci\'on y de Estudios Avanzados del IPN, Unidad M\'erida\\ 
A.P. 73, Cordemex. M\'erida, Yucat\'an 97310. MEXICO\\ 
}
\author{ 
G.~S\'anchez-Col\'on\footnote{
Permanent Address:
Departamento de F\'{\i}sica Aplicada.
Centro de Investigaci\'on y de Estudios Avanzados del IPN, Unidad M\'erida.
A.P. 73, Cordemex. M\'erida, Yucat\'an 97310. MEXICO.
}
}
\address{
Physics Department.
University of California\\
Riverside, California 92521-0413. U.S.A.\\
}

\date{May 11, 1998}

\maketitle

\begin{abstract}
Strong-flavor and parity a priori mixing in hadrons are shown to describe 
well the experimental evidence on weak radiative decays of hyperons.
An independent determination of the a priori mixing angles is performed.
The values obtained for them are seen to have a universality--like property,
when compared to their values in non-leptonic decays of hyperons.
\end{abstract}


The enhancement phenomenon observed in weak decays of hadrons has remained 
an open challenge for quite sometime now. In a recent paper~\cite{LE-6129} we
have studied the possibility of attributing this phenomenon to the existence
of a priori strong-flavor and parity mixings in hadrons.
In non-leptonic decays of hyperons (NLDH) we obtained two important results:
(i) the predictions of the $|\Delta I| = 1/2$ rule and (ii) a fairly good
description of the available experimental data.
The a priori mixing angles, three in the case of NLDH which we named
$\sigma$, $\delta$, and $\delta'$, appear as free parameters because we did not
have any theoretical or phenomenological control on them.
We obtained the values
$\sigma = (4.9 \pm 1.5)\times 10^{-6}$,
$|\delta| = (0.23\pm 0.07)\times 10^{-6}$,
and
$|\delta'| = (0.26\pm 0.07)\times 10^{-6}$.
A more detailed analysis~\cite{nldh-d} indicates that the violations of the
$|\Delta I| = 1/2$ rule affect their values and one should take more
conservative ones as their estimates in NLDH, namely,
$\sigma = (4.9 \pm 2.0)\times 10^{-6}$,
$|\delta|= (0.22 \pm 0.09)\times 10^{-6}$,
and
$|\delta'| = (0.26 \pm 0.09)\times 10^{-6}$.
The explicit absolute-values bars on $\delta$ and $\delta'$ are introduced as
a reminder that their signs were not determined in NLDH.

The results obtained so far are very encouraging.
Indeed, the a priori mixing scheme may serve as framework for the systematic
description of the enhancement phenomenon.
However, along this line, one faces very stringent tests that this scheme
must pass successfully.
Since the a priori mixings are non-perturbative effects in hadron wave
functions, not only other decays processes originating from different dynamics
where enhancement is present should be well described but the a priori angles
$\sigma$, $\delta$, and $\delta'$ should appear with values compatible with
their values in NLDH, i.e., these angles must necessarily have a
universality-like property.
If this last is not obtained, then this should be sufficient reason to reject
the whole scheme, at least as one with practical usefulness.

In this letter we shall make a detailed application of the above scheme to the 
weak radiative decays of hyperon (WRDH).
The experimental evidence~\cite{pdg} on them is collected in column eight of
Table~\ref{tablai}.
If the enhancement phenomenon describes overwhelmingly these decays and if
such phenomenon is described by a priori mixings, then these transition are
due to the matrix elements of the ordinary parity and flavor conserving
electromagnetic interaction hamiltonian
$H_{em} = iJ^\mu_{em}A_\mu$
between a priori mixed baryons.
The latter are given by 

\[
p_{ph} = 
p_{0s} - \sigma \Sigma^+_{0s} - \delta \Sigma^+_{0p}
+ \cdots
,
\]
 
\[
n_{ph} = 
n_{0s} +
\sigma ( \frac{1}{\sqrt{2}} \Sigma^0_{0s} + \sqrt{\frac{3}{2}} \Lambda_{0s}) +
\delta ( \frac{1}{\sqrt{2}} \Sigma^0_{0p} + \sqrt{\frac{3}{2}} \Lambda_{0p} )
+ \cdots
,
\]
             
\[
\Sigma^+_{ph} =
\Sigma^+_{0s} + \sigma p_{0s} - \delta' p_{0p}
+ \cdots
,
\]
 
\begin{equation}
\Sigma^0_{ph} =
\Sigma^0_{0s} +
\frac{1}{\sqrt{2}} \sigma ( \Xi^0_{0s}- n_{0s} ) +
\frac{1}{\sqrt{2}} \delta \Xi^0_{0p} + \frac{1}{\sqrt{2}} \delta' n_{0p}
+ \cdots
,
\label{bph}
\end{equation}
 
\[
\Sigma^-_{ph} = \Sigma^-_{0s} + \sigma \Xi^-_{0s} + \delta \Xi^-_{0p} 
+ \cdots
,
\]

\[
\Lambda_{ph} = 
\Lambda_{0s} + 
\sqrt{\frac{3}{2}} \sigma ( \Xi^0_{0s}- n_{0s} ) +
\sqrt{\frac{3}{2}} \delta \Xi^0_{0p} + 
\sqrt{\frac{3}{2}} \delta' n_{0p}
+ \cdots
,
\]
 
\[
\Xi^0_{ph} =
\Xi^0_{0s} -
\sigma
( \frac{1}{\sqrt{2}} \Sigma^0_{0s} + \sqrt{\frac{3}{2}} \Lambda_{0s} ) +
\delta'
( \frac{1}{\sqrt{2}} \Sigma^0_{0p} + \sqrt{\frac{3}{2}} \Lambda_{0p} )
+ \cdots
,
\]

\[
\Xi^-_{ph} =
\Xi^-_{0s} - \sigma \Sigma^-_{0s} + \delta' \Sigma^-_{0p}
+ \cdots
.
\]

\noindent
In these expressions the subindices naught, $s$, and $p$ indicate
strong-flavor, positive, and negative parity eigenstates and each physical
baryon is the mass eigenstate already observed~\cite{pdg}.
$A_\mu$ represents the photon field operator and $J_\mu$ is the ordinary
electromagnetic current operator.
This operator is --- and this must be emphasized --- a flavor-conserving
Lorentz proper four-vector.
The hadronic parts of the transition amplitudes are then given by:

\[
\langle p_{ph} | J^{\mu}_{em} | \Sigma^+_{ph} \rangle =
\bar u_p [ \sigma ( f^{\Sigma^+}_2 - f^p_2 ) +
           (\delta'f^p_2 - \delta f^{\Sigma^+}_2) \gamma^5 ]
i\sigma^{\mu\nu}q_{\nu} u_{\Sigma^+}
\]

\[
\langle \Sigma^-_{ph} | J^{\mu}_{em} | \Xi^-_{ph} \rangle =
\bar u_{\Sigma^-} [ \sigma ( f^{\Xi^-}_2 - f^{\Sigma^-}_2 ) +
                    (\delta' f^{\Sigma^-}_2 - \delta f^{\Xi^-}_2) \gamma^5 ] 
i\sigma^{\mu\nu}q_{\nu} u_{\Xi^-}       
\]

\begin{eqnarray}
\langle n_{ph} | J^{\mu}_{em} | \Lambda_{ph} \rangle 
& = &
\bar u_n 
\left\{ 
\sigma \left[ \sqrt{\frac{3}{2}} ( f^{\Lambda}_2 - f^n_2 ) + 
                   \frac{1}{\sqrt 2} f^{\Sigma^0\Lambda}_2 \right]
\right.
\nonumber \\
& &
\left.
+ \left[
\sqrt{\frac{3}{2}} 
(\delta' f^n_2  - \delta f^{\Lambda}_2) 
- 
\delta \frac{1}{\sqrt 2} f^{\Sigma^0\Lambda}_2 
\right]
\gamma^5 
\right\}
i\sigma^{\mu\nu}q_{\nu} u_{\Lambda} 
\nonumber
\end{eqnarray}  

\begin{eqnarray}
\langle \Lambda_{ph} | J^{\mu}_{em} | \Xi^0_{ph} \rangle 
& = &
\bar u_{\Lambda} 
\left\{ 
\sigma \left[ \sqrt{\frac{3}{2}} ( f^{\Xi^0}_2 - f^{\Lambda}_2 ) -
         \frac{1}{\sqrt 2} f^{\Sigma^0\Lambda}_2 \right]
\right.
\nonumber \\
& &
\left.
+ \left[
\sqrt{\frac{3}{2}} (\delta' f^{\Lambda}_2 - \delta f^{\Xi^0}_2 ) 
+ 
\delta' \frac{1}{\sqrt 2} f^{\Sigma^0\Lambda}_2 
\right] 
\gamma^5 
\right\}
i\sigma^{\mu\nu}q_{\nu} u_{\Xi^0}
\nonumber       
\end{eqnarray}

\begin{eqnarray}
\langle \Sigma^0_{ph} | J^{\mu}_{em} | \Xi^0_{ph} \rangle 
& = &
\bar u_{\Sigma^0} 
\left\{ 
\sigma \left[ \frac{1}{\sqrt 2} ( f^{\Xi^0}_2 - f^{\Sigma^0}_2 ) - 
         \sqrt{\frac{3}{2}} f^{\Sigma^0\Lambda}_2 \right]
\right.
\nonumber \\
& &
\left.
+ \left[
\frac{1}{\sqrt 2} (\delta' f^{\Sigma^0}_2 - \delta f^{\Xi^0}_2 )               
+ 
\delta' \sqrt{\frac{3}{2}}f^{\Sigma^0\Lambda}_2  
\right]
\gamma^5 
\right\}
i\sigma^{\mu\nu}q_{\nu} u_{\Xi^0}  
\label{seis} 
\end{eqnarray}

\noindent
In these amplitudes only contributions to first order in $\sigma$, $\delta$,
and $\delta'$ need be kept.
Each matrix element is flavor and parity conserving and can be expanded in
terms of charge $f_1(0)$ form factors and anomalous magnetic $f_2(0)$ form
factors.
Because the charges of the positive- and negative-parity parts of the same
physical wave function are equal and such changes are controlled by the
generator property of $J_\mu$ all the $f_1$'s cancel away and only the $f_2$
contribute. 
The $f_2$ between $s$ and $p$ parts and between $p$ and $s$ parts can be 
identified with the $f_2$ between $s$ and $s$ parts, provided that a relative 
minus sign be present between the former two in order to respect hermicity.
Notice that the amplitudes~(\ref{seis}) are all of the form
$\bar{u}_B(C+D\gamma_5)i\sigma^{\mu\nu}q_{\nu}u_A$, where $C$ is the so-called
parity conserving amplitude and $D$ is the so-called parity violating one.
We stress, however, that in this model both $C$ and $D$ are parity and flavor
conserving.

In principle, we have information about all the quantities that appear in
these amplitudes for WRDH.
The a priori angles are known from NLDH and the $f_2$ can be related to the
measured total magnetic moments of spin 1/2 baryons.
The latter values are displayed in the second column of
Table~\ref{tablai}~\cite{pdg}.
However, it is important that the mixing angles be determined independently in
WRDH and, accordingly, we shall use them as free parameters in the remaining
of this paper.
How the $f_2$'s are related to the observed total magnetic moments is a 
question we shall deal with in steps.
As a first approximation we shall assume that the $f_2$'s are related to the
$\mu$'s of Table~\ref{tablai} by the formula $\mu^{exp}_A = e_A + f^A_2$ where
$A$ is a baryon and $e_A$ its charge.
Thus, for example, $f^p_2$ obeys the relationship $\mu^{exp}_p = 1 + f^p_2$
(in nuclear magnetons), etc. Using this assumption we may compare with the
experimental data of WRDH.
The predictions obtained are displayed in the columns~I of
Table~\ref{tablai}.

These first results are not quite good yet but they have a qualitative value. 
The main point is that the a priori mixing angles come out with the same order 
of magnitude observed in NLDH, which is very encouraging.
The predictions for the observables are some very good, some good, but some
show important deviations.
The latter still have qualitative value, but should be improved. 
The  values of the $\mu$'s agree fairly well with their experimental 
counterparts.

As an intermediate step in this analysis it turns out to be very helpful to 
see what are the values of the total magnetic moments required to reproduce
well the experimental observables of WRDH.
This is achieved by relaxing the error bars of the measured $\mu$'s up to
10\% of the corresponding central values and, then, repeating the previous
step.
The results are displayed in the columns~II of Table~\ref{tablai}.
The experimental data are very well reproduced now, but at the expense of
sizable (several percent) changes in the $\mu$'s and new values for the mixing
angles.

This second step clearly shows that we must accept that our first 
approximation ---that of identifying the experimentally measured $\mu$'s with 
the ones that are actually related to the $f_2$'s in this approach to
WRDH--- must be improved.
The $\mu$'s to be used for determining the $f_2$'s in the WRDH amplitudes are
really transition magnetic moments.
For example, the measured value of $\mu_p$ corresponds to the matrix element
$\langle p_{ph}|J^\mu_{em}|p_{hp}\rangle \simeq
\langle p_{os}|J^\mu_{em}|p_{os}\rangle$,
where both physical wave functions carry the mass $m_p$.
In contrast, the $\mu_p$ the appears in $\Sigma^+ \to p\gamma$ corresponds to
a matrix element whose bra carries the mass $m_p$ and whose ket carries the
mass $m_{\Sigma^+}$.
So, the normalization of $\mu_p$ originating in the matrix element
$\langle p_{ph}|J_\mu |\Sigma^+_{ph}\rangle$
should be related to both masses, $m_p$ and $m_{\Sigma^+}$.
It is in this sense that the magnetic moments that we must use are transition
magnetic moments.

The natural normalization of magnetic moments is determined by the Gordon 
decomposition~\cite{bjorken}.
Using this expansion for guidance, then, for example, $\mu_p$ should be
normalized to $m_p + m_{\Sigma^+}$ and not to $2m_p$, etc. 
One can see already a qualitative indication of this happening in the first 
column~II in Table~\ref{tablai}, the changes in the $\mu$'s are
systematically in this direction.
$\mu_p$, $\mu_n$, $\mu_{\Xi^-}$, $\mu_{\Sigma^-}$, 
$\mu_{\Sigma^+}$, $\mu_{\Xi^0}$, and $\mu_{\Sigma^0}$ appear to become 
smaller or larger according to such changes in normalization.
$\mu_\Lambda$ and $\mu_{\Sigma^0\Lambda}$ are mixed cases because they appear
in two or three decays and will be required to be reduced or to be increased
in going from one case to another and, therefore, Table~\ref{tablai} cannot
provide a clear cut tendency.

Our third step is to improve our approximation following the above discussion. 
One must change the normalization of the total magnetic moments either by 
applying, for example, the factor $(m_p+m_p)/(m_p+m_{\Sigma^+})$ to the
experimental $\mu_p$ or the inverse factor to the theoretical $\mu_p$ related
to $f^p_2$. 
Numerically, either way leads to the same result.
For definiteness, we choose the former.
The corrected experimental values are displayed in column five of
Table~\ref{tablai}.
Then, recalculating everything lead to the predictions of columns labeled~III
of Table~\ref{tablai}.
The values of the mixing angles appear in the bottom of the last column of
this table.

The overall agreement is greatly improved, the experimental data are well 
produced while keeping the magnetic moments in very good agreement with their 
experimental counterparts.
The only deviations that merit further discussion appear in $\Gamma_2$ and in
$\mu_{\Sigma^-}$ which are intimately related. 
These deviations are probably due to another one of our approximations: 
ignoring the contributions of $W_\mu$~\cite{bassaleck}.
We have assumed these contributions to be small (non-enhanced) and we have
neglected them, but since they are non-zero they should eventually be included.
Their effect should appear mainly in small amplitudes.
$\Gamma_2$ is about an order of magnitude smaller than other $\Gamma$'s and,
accordingly, its corresponding amplitudes are also the smallest ones.
The interference of the $W_\mu$ amplitudes should be relatively more important
in this case.
Therefore, one is not entitled to expect a better agreement of a priori
mixings in this case.
The agreement already obtained is probably the best one can hope for if one
stays short of actually incorporating the contributions of $W_\mu$, as we
have.

In Table~\ref{tablai} we included the predictions for $\alpha_2$ and
$\alpha_3$.
They are also quite sensible to the a priori mixing parameters.
It would be important to have experimental information on them.
At present only $\alpha_2=1.0\pm1.3$ is available~\cite{lach}.
This measurement can be understood to mean that only its sign is known, namely,
positive.
The prediction in case III agrees with this sign.

We are now in a position to conclude our present analysis.
To extend the credibility of the a priori mixing scheme it was very important
to be able to describe WRDH.
As we have shown above this is achieved.
However, the most important and stringent test is that the mixing angles share
a universality-like property.
The values for them obtained independently in WRDH are in very good agreement
with the absolute values obtained for them in NLDH.
It is the passing this universality-like test that lends the strongest 
support to the possibility that the above scheme may serve a framework for
the systematic description of the enhancement phenomenon in weak decays of 
hadrons.

As a closing remark, let us mention how the three orders of magnitudes 
observed experimentally between NLDH and WRDH can be understood: The 
enhancement phenomenon dominates these decays, such phenomenon is described by 
the a priori mixings scheme, and the mixing angles are common to both dynamics, 
then, the relative order of magnitude between these two sets of decays is 
given by the ratio $(g^2/4\pi)/(e^2/4\pi) \sim 10^3$, where $g$ is the 
strong-interaction Yukawa coupling constant\cite{dumbrajs}.

The authors wish to acknowledge partial support by CONACyT (M\'exico).

\begin{table}
\squeezetable
\caption{
Experimental values of the observables in WRDH and of the magnetic moments
(m.\ m.) of hyperons along with the predictions of the three cases considered.
The indeces 1,2,...,5 on the observables
correspond, respectively, to the decays
$\Sigma^+\to p\gamma$, $\Xi^-\to\Sigma^-\gamma$, $\Lambda\to n\gamma$,
$\Xi^0\to\Lambda\gamma$, and $\Xi^0\to\Sigma^0\gamma$.
The numbers in parenthesis in the m.\ m.\ indicate the decay in which they appear.
The values of the a priori mixing angles in column eight come from NLDH. 
The mixing angles are in 10$^{-6}$, the decay rates are in $10^6$~sec$^{-1}$, 
and the m.\ m.\ are in nuclear magnetons.
The only m.\ m.\ that has not been measured is $\mu_{\Sigma^0}$.
We have taken for it its $SU(3)$ estimate with a 10\% error bar.
}~
\label{tablai}
\begin{tabular}
{
l
r@{.}l@{$\pm$}r@{.}l
d
d
r@{.}l@{$\pm$}r@{.}l
d
c
r@{.}l@{$\pm$}r@{.}l
d
d
d
}
&
\multicolumn{6}{c}{Magnetic moments} &
\multicolumn{5}{c}{Transition m.\ m.} &
&
\multicolumn{7}{c}{Observables and angles}
\\
&
\multicolumn{4}{c}{Exp.} &
I &
II &
\multicolumn{4}{c}{Exp.} &
III &
&
\multicolumn{4}{c}{Exp.} &
I &
II &
III
\\
\tableline
$\mu_p (1)$ &
2 & \multicolumn{3}{l}{793} &
2.793 &
2.745 &
2 & \multicolumn{3}{l}{463} &
2.463 &
$\Gamma_1$ &
15 & 65 & 0 & 88 &
11.55 &
15.60 &
14.62
\\
$\mu_n (3)$ &
$-$1 & \multicolumn{3}{l}{913} &
$-$1.913 &
$-$1.654 &
$-$1 & \multicolumn{3}{l}{750} &
$-$1.750 &
$\Gamma_2$ &
0 & 77 & 0 & 14 &
1.37 &
0.81 &
1.30
\\
$\mu_{\Xi^-} (2)$ &
$-$0 & 651 & 0 & 003 &
$-$0.652 &
$-$0.747 &
$-$0 & 683 & 0 & 003 &
$-$0.685 &
$\Gamma_3$ &
6 & 65 & 0 & 57 &
7.13 &
6.67 &
6.16
\\
$\mu_{\Sigma^-} (2)$ &
$-$1 & 160 & 0 & 025 &
$-$1.018 &
$-$0.868 &
$-$1 & 103 & 0 & 024 &
$-$0.958 &
$\Gamma_4$ &
3 & 66 & 0 & 56 &
5.26 &
3.68 &
4.38
\\
$\mu_\Lambda (3)$ &
$-$0 & 613 & 0 & 004 &
$-$0.611 &
$-$0.553 &
$-$0 & 665 & 0 & 004 &
$-$0.665 &
$\Gamma_5$ &
12 & 1 & 1 & 4 &
5.23 &
11.8 &
10.13
\\
$\mu_\Lambda' (4)$ &
\multicolumn{4}{c}{} &
&
&
$-$0 & 563 & 0 & 004 &
$-$0.562 &
$\alpha_1$ &
$-$0 & 76 & 0 & 08 &
$-$0.78 &
$-$0.75 &
$-$0.88
\\
$\mu_{\Sigma^+} (1)$ &
2 & 458 & 0 & 010 &
2.430 &
2.553 &
2 & 748 & 0 & 011 &
2.763 &
$\alpha_2$ &
\multicolumn{4}{c}{} &
$-$0.09 &
0.56 &
0.20
\\
$\mu_{\Xi^0} (4)$ &
$-$1 & 250 & 0 & 014 &
$-$1.271 &
$-$1.502 &
$-$1 & 353 & 0 & 015 &
$-$1.357 &
$\alpha_3$ &
\multicolumn{4}{c}{} &
$-$0.79 &
$-$0.83 &
0.76
\\
$\mu_{\Xi^0}' (5)$ &
\multicolumn{4}{c}{} &
&
&
$-$1 & 311 & 0 & 015 &
$-$1.305 &
$\alpha_4$ &
0 & 40 & 0 & 40 &
$-$0.86 &
0.18 &
0.27
\\
$\mu_{\Sigma^0\Lambda} (3)$ &
$-$1 & 610 & 0 & 080 &
$-$1.597 &
$-$1.626 &
$-$1 & 808 & 0 & 090 &
$-$1.659 &
$\alpha_5$ &
0 & 20 & 0 & 32 &
$-$0.46 &
$-$0.02 &
0.77
\\
$\mu_{\Sigma^0\Lambda}' (4)$ &
\multicolumn{4}{c}{} &
&
&
$-$1 & 529 & 0 & 076 &
$-$1.426 &
$\sigma$ &
4 & 9 & 2 & 0 &
\multicolumn{1}{r}{0.70$\pm$0.03} &
\multicolumn{1}{r}{1.72$\pm$0.46} &
\multicolumn{1}{r}{0.91$\pm$0.06}
\\
$\mu_{\Sigma^0\Lambda}'' (5)$ &
\multicolumn{4}{c}{} &
&
&
$-$1 & 482 & 0 & 074 &
$-$1.617 &
$\delta$ & 
$|$0 & 22$|$ & 0 & 09 &
\multicolumn{1}{r}{0.03$\pm$0.08} &
\multicolumn{1}{r}{$-$0.40$\pm$0.19} &
\multicolumn{1}{r}{$-$0.11$\pm$0.08}
\\
$\mu_{\Sigma^0} (5)$ &
0 & 649 & 0 & 065 &
0.499 &
0.624 &
0 & 617 & 0 & 062 &
0.500 &
$\delta'$ &
$|$0 & 26$|$ & 0 & 09 &
\multicolumn{1}{r}{0.10$\pm$0.07} &
\multicolumn{1}{r}{$-$0.28$\pm$0.18} &
\multicolumn{1}{r}{$-$0.25$\pm$0.08}
\\
\end{tabular}
\end{table}

\end{document}